\documentclass[conference]{IEEEtran}
\IEEEoverridecommandlockouts
\usepackage{cite}
\usepackage{amsmath,amssymb,amsfonts}

\usepackage{graphicx}
\usepackage{textcomp}
\usepackage{xcolor}
\usepackage{lipsum}
\usepackage[noend]{algpseudocode}
\usepackage{makecell}
\usepackage{algorithm}

\usepackage{algpseudocode}
\makeatletter

\def\BibTeX{{\rm B\kern-.05em{\sc i\kern-.025em b}\kern-.08em
    T\kern-.1667em\lower.7ex\hbox{E}\kern-.125emX}}
\makeatletter
\newcommand*{\rom}[1]{\expandafter\@slowromancap\romannumeral #1@}

\makeatother
\begin{document}

\title{Active Learning Solution on Distributed Edge Computing\\
\thanks{The research leading to these results has received funding from the European Union’s Horizon 2020 research and innovation program under the Marie Sklodowska-Curie grant agreement No. 764785, FORA-Fog Computing for Robotics and Industrial Automation. This work was further supported by the Danish Innovation Foundation through the Danish Center for Big Data Analytics and Innovation (DABAI).}
}
\author{\IEEEauthorblockN{1\textsuperscript{st} Jia Qian, Sayantan Sengupta,  Lars Kai Hansen}
\IEEEauthorblockA{\textit{Department of Applied Mathematics and Computer Science} \\
\textit{Technical University of Denmark}\\
Lyngby, Denmark \\
\{jiaq, says, lkai\}@dtu.dk}

}

\maketitle

\begin{abstract}
Industry 4.0 becomes possible through the convergence between Operational and Information Technologies. All the requirements to realize the convergence is integrated on the Fog Platform.
Fog Platform is introduced between cloud server and edge devices when the unprecedented generation of data causes the burden of cloud server, leading the ineligible latency. In this new paradigm, we divide  the computation tasks and push it down to edge devices. Furthermore, local computing (at edge side) may improve privacy and trust. To address these problems, we present a new method, in which we decompose the data aggregation and processing, by dividing them between edge devices and fog nodes intelligently. We apply active learning on edge devices; and federated learning on the fog node which significantly reduces the data samples to train the model as well as the communication cost. To show the effectiveness of the proposed method, we implemented and evaluated its performance for an image classification task. In addition, we consider two settings: massively distributed and non massively distributed and offer the corresponding solutions.
\end{abstract}


\section{Introduction}
We are at the beginning of a new industrial revolution - Industry 4.0 - which will bring increased productivity and flexibility, mass customization, reduced time-to-market, improved product quality, innovations, and new business models. However, Industry 4.0 will only become a reality through the convergence of Operational and Information Technologies (OT and IT), which are currently separated in a hierarchical pyramid and use different computation and communication technologies. OT related applications are typically safety-critical and real-time (RT), which means requiring, guaranteed extra-functional properties as real-time behavior, reliability, availability, industry-specific safety standards, and security.

IT such as Cloud Computing and Service Oriented Architecture cannot offer the stringent properties
required from the lower level OT, though, IT is indeed present in the upper level. The combination
of IP-protocols, the interoperability standards and other techniques may form a Fog Computing
platform, which enables the computation, communication and storage closer to the network. Essentially, it is a
 shift from the cloud-based paradigm towards a fog computing paradigm. In cloud-based paradigm, the devices send all the information to a centralized authority, which processes the data. However, this paradigm introduces a considerable latency, not all the application may tolerate it, and also the computation task is becoming heavier for the remote server. In Fog Computing (FC, \cite{b1}) or Edge Computing (EC) the data processing computation will be distributed among edge devices, fog devices, and the cloud server. FC and EC are interchangeable in the following text. This emergence of EC is mainly in response to the heavy workload at the cloud side and the significant latency at the user side. To reduce the delay in fog computing, the concept of fog node is introduced. The Fog Node (FN) \cite{b2} is essentially a platform placed between Cloud and Edge Devices (ED) as middleware, and it will further facilitate the 'things' to realize their potentials \cite{b3}. This change of paradigm will help application domains, such as industrial automation, robotics or autonomous vehicles, where real-time decision making by using machine learning approaches is crucial.
 
 

Federated Learning \cite{b6} allows the centralized server training models without moving the data to a central location. In particular, FL is used in a distributed configuration, in which the model is built without direct access to training data, and indeed the data remains in its generation location, which provides both data locality and privacy. In the beginning, a server coordinates a set of nodes, each with training data that cannot be accessed by server directly. These nodes train a local model and share individual models with the server. The server uses these individual models to create a federated model and sends the model back to the nodes. Then, another round of local training takes place, and the process continues. Nevertheless, this extra work on edge devices has to be minimized by selecting the most important data samples needed to build the local model. In this context, we want to use Active Learning (AL) as a more effective learning framework. AL chooses training data efficiently when labeling data becomes drastically expensive.


Motivated by the above mentioned, the research issues and possible direction, we propose a new scheme. In literature, there exist some papers that discuss the application of machine-learning algorithm directly on the Fog node platform \cite{b32}\cite{b21}. As we have already discussed, to efficiently use fog infrastructure, we need to delegate the work among fog node and edge devices. Hence, in this paper, we propose a new efficient privacy-preserving data analytic scheme in Fog environment. We offer using federated learning at the centralized fog nodes to create the initial training model. To improve the performance further, we recommend using Monte-Carlo dropout Bayesian Neural Network to quantify the model uncertainty to apply AL at the edge devices, by selecting the sample points effectively. All in all, we propose a possible solution in Edge Computing setting, where the user privacy, training cost, and upload bottleneck are the main issues to address. Our strategy may reduce the training cost by applying AL and preserve the user privacy and reduce the communication by using FL. Moreover, the proof of concept is demonstrated by applying the method on a benchmark dataset.

The remainder of this paper is organized as follows.
Section{~\ref{sec:related}} explains preliminary concepts and the related work, in particular, LeNet model, FL, Monte-Carlo dropout Bayesian Neural Network, AL framework and so on. In Section~\ref{sec:ourapproach}, we introduce the proposed scheme. Section~\ref{sec:eval} covers the details of our experimental design and data collection strategy followed by a discussion on our results. Section~\ref{sec:Conc} concludes the paper. %

\section{Preliminary Concepts and Related Work}
\label{sec:related}
In this section, we discuss the different techniques that are essential for our scheme. We also present a brief overview of the major research studies related to our work.
\subsection{LeNet}
LeNet \cite{b40} is classic Convolutional Neural Network, proposed by Yann LeCun for the first time, which is composed by three convolutional layers, two pooling layers and the fully-connected layer to compress the convolutional layers. The specific architectures is described in Table \ref{tab2}, and we will use it as our basic Neural Network model.

\begin{table}[htbp]
\caption{LeNet architecture}
\begin{center}
\begin{tabular}{|c|c|c|c|}
\hline
 \textbf{layer}&\textbf{layer name} & \makecell{output channels or \\number of nodes} & \textbf{kernel size}\\
\hline
\textbf{\makecell{1}} & \textbf{Convolution} & 6 &5x5\\
\hline
\textbf{\makecell{2}} & \textbf{Ave. Pooling} & 6&2x2\\
\hline
\textbf{\makecell{3}}& \textbf{Convolution}  & 16&5x5\\
\hline
\textbf{\makecell{4}}&\textbf{Ave. Pooling} & 16 &2x2 \\
\hline
\textbf{\makecell{5}} &\textbf{Convolution}& 120&5x5\\
\hline
\textbf{\makecell{6}} &\textbf{FC}& 84 &-\\
\hline
\textbf{\makecell{output}} &\textbf{FC}& 10 &-\\
\hline
\end{tabular}
\label{tab2}
\end{center}
\end{table}

\subsection{Federated Learning}
Federated learning (FL) is a collaborative form of machine learning where the training process is distributed among many users; this enables to build machine learning systems without complete access to training data \cite{b6}. In FL, the data remains in its birth location, which helps to ensure privacy and reduces communication costs. In principle, this idea can be applied to any model for which the criterion of updates can be defined, which naturally includes the methods based on gradient descent, which nowadays most of the popular models do. For instance, linear regression, logistic regression, neural networks, and linear support vector machines can all be used for FL by letting users compute gradients \cite{b35} \cite{b36}.


We define the goal of FL as learning a model with parameters embodied in matrix from data stored across a large number of clients (Edge Devices). Suppose the server (FN) distributes the model (at round t) $W_t$ to N clients for further updating, and the updated models are denoted as $W_t^1, W_t^2, W_t^3,...W_t^N$. Then, the clients send the updated models back to the server, and the server updates the model W according to the aggregated information.
\begin{equation}
 \begin{split}
W_{t+1} := \sum_i^N \alpha_{i} * W_t^i
 \end{split}
\end{equation}
Where $\alpha$ can be uniformly distributed or according to the $t-1$ round performance, we use the former one in our work, namely, average the parameters. The learning process can be iteratively carried out.
\subsection{Active Learning}
Active learning (AL) is a particular case of machine learning in which a learning algorithm interactively queries the user to obtain the desired outputs at new data points. Typically, AL achieves higher performance with the same number of data samples, or achieves a given performance using less data. Active learning is an appropriate choice when i) labeling data is expensive, or ii) limited data collection. Initially, the researchers fit the machine learning algorithms that mostly work for tabular data to the active learning framework. Recently, it starts registering with the deep neural network, though, it seemingly contradicts with each other as deep neural network typically requires large training data.

\begin{figure}
\centerline{\includegraphics[width=70mm,scale=0.5]{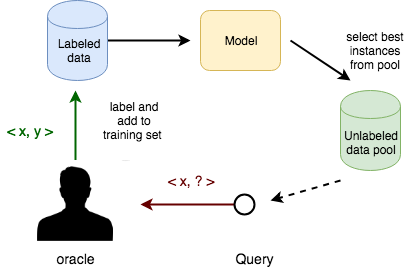}}
\caption{Pool-based Active Learning Framework.}
\label{pool_framework}
\end{figure}

Active learning can be divided into two categories: pool-based and stream-based. Stream-based active learning typically draws one at a time from the input source, and the learner must decide whether to query or discard it, whereas pool-based queries the most informative instance from a large pool (See Fig \ref{pool_framework}). We consider the pool-based scheme: the model chooses the "good" instances from the unlabeled pool according to the acquisition function and then ask the Oracle to label them. Sequentially, we include the labeled ones to the training set for the further training. We can repeat such operations for several times. The authors in \cite{b14} showed that a pool-based support vector machine classifier significantly reduces the needed data to reach some particular level of accuracy in text classification, analogously, for image retrieval application in \cite{b20}.

\subsection{Acquisition Function}\label{dr}
The acquisition function is a measure of how desirable the given data point is, either considering minimizing the loss or maximizing the likelihood. Uncertainty-based methods aim to use uncertain information to enhance the model during the training process. It plays the role of the exploitation while acts as the exploration part. We will introduce three different ways to estimate the uncertainty.


\textbf{-- Maximal Entropy:} $H[y|x,D_{train}]$ is the predictive entropy expectation as defined in \cite{b10}.
\begin{equation}
 \begin{split}
H[y|x,D_{train}] := -\sum_{c}(\frac{1}{T}\sum_{t=1}^{T}p(y=c|x,w_{t}))*\\
 \log (\frac{1}{T}\sum_{t=1}^{T}p(y=c|x,w_{t}))\label{al}
 \end{split}
\end{equation}

\textbf{-- BALD:} (Bayesian Active Learning by Disagreement\cite{b16}) it measures the mutual information between data and weights, and it can be interpreted as seeking data points for which the parameters (weights) under the posterior disagrees the most.
\begin{equation}
 \begin{split}
I[y;w|x,D_{train}]&:= H[y|x,D_{train}]-\frac{1}{T}\sum_{t}\sum_{c}\\
    &-p(y=c|x,w_{t})\log p(y=c|x,w_{t})
 \end{split}
\end{equation}
\textbf{-- Variational Ratios (VR):} it maximises the variational ratio by considering the followings\cite{b17}:
\begin{equation}
 \begin{split}
    V[x]:= 1- \max_{y} p(y|x, D_{train})
 \end{split}
\end{equation}
It is similar to Maximal Entropy, but less effective as reported in \cite{b18}.

\section{Proposed scheme}\label{sec:ourapproach}
The above-mentioned acquisition functions are based on the quantification of the uncertainty of the model, thus we need a probabilistic model to address it. We will introduce a vital concept, Bayesian neural network, which is the foundation that AL can work appropriately on image classification by using neural network.

For the probabilistic model, learning the posterior distribution is identical to the training process for the deterministic model. We define it as the following according to the Bayes Equation, expressed by the likelihood $p(D|w)$ and prior distribution $p(w)$. $D$ is the dataset, $w$ is the model parameters.
\begin{equation}
    p(w|D) = \frac{p(D|w)p(w)}{p(D)}
\end{equation}
Similarly, the predictive posterior distribution is identical to the prediction. $x^*$ is the new input to predict. Rather than just outputting a single prediction, the probabilistic model outputs a normal distribution, where the mean is the most likely prediction.
\begin{equation}
    p(y^*|x^*,D) = \int p(y^*|x^*,w)p(w|D) \mathrm{d}w
\end{equation}
\subsection{Bayesian Neural Network approximating by Dropout}\label{bnn}

In neural network, the posterior is intractable \cite{b37} (no closed analytical form), we need a variational distribution $q(w)$ to approximate it. Thus, we use KL divergence to measure the distance between $q(w)$ and the real posterior distribution $p(w|D)$, aiming to minimize it.
\begin{equation}\label{eq7}
\begin{split}
       D_{KL}(q(w)||p(w|D))=\int q(w)\log \frac{q(w)}{p(w|D)}\mathrm{d}w\\=\int q(w)\log \frac{q(w)p(D)}{p(w|D)p(D)}\mathrm{d}w\\=\int q(w)\log \frac{q(w)p(D)}{p(w,D)}\mathrm{d}w\\=E_{q(w)}[\log \frac{q(w)p(D)}{p(D,w)}]\\=E_{q(w)}[\log q(w)-\log p(w,D)]+\log p(D)\geqslant 0
\end{split}
\end{equation}
Since KL distance $D_{KL}$ is greater equal than zero, we can derive equation \ref{eq7} as follows:
\begin{equation}
\begin{split}
    \log p(D) \geqslant E_{q(w)}[\log p(w,D)-\log q(w)]\\= E_{q(w)}[\log{p(D|w)p(w)}-\log q(w)]\\=E_{q(w)}[\log p(D|w)+\log p(w)-\log q(w)]\\=E_{q(w)}[\log p(D|w)-\log \frac{q(w)}{p(w)}]\\=E_{q(w)}[\log p(D|w)]-D_{KL}(q(w)||p(w))
\end{split}
\end{equation}
Now we can define the ELBO (Evidence Lower Bound), composed of two parts, the reconstruction error and KL distance between variational prior distribution and true prior distribution. Notably, after the mathematical manipulation, we convert the KL distance between posterior distribution to prior distribution. 
\begin{equation}\label{eq9}
    \mathcal{L}_{VI}:= E_{q(w)}[\log p(D|w)]-KL(q(w)||p(w))
\end{equation}


Bayesian Neural Network distinguishes to the normal Neural Network by placing a prior distribution over the weights of the model. Let's define the weights of neurons in layer $i$ as $W = (w_i)_{i=1}^L$ and the variational prior distribution $q(w)$ as a Bernoulli distribution.
\begin{equation}
    \centering
    W_i = M_i * \text{diag}([Z_{i,j}]_{j=1}^{K_i}),
\end{equation}
where
\begin{equation}
    Z_{i,j} \thicksim Bernoulli (p_i, dropout),  i = 1,...,L, j =1,..K_{i-1}
\end{equation}
$M_i$ are variational parameters to be optimized. The diag(.) maps vector to diagonal matrices, whose diagonal are the elements of the vector. $K_i$ indicates the number of neurons in layer i, L is the number of layers.  Then we can re-write the cost function in equation \ref{eq9} as $\hat{\mathcal{L}}_{VI}$:
\begin{equation}
    \hat{\mathcal{L}}_{VI}:= \sum_{i=1}^N E(y_i,\hat{f}(x_i,\hat{W}_i))-KL(q(w)||p(w))
\end{equation}
Where $\widehat{w}_t \thicksim q(w)$ and $\hat{f}(.)$ is the neural network that outputs the prediction given the parameters $\hat{W}_i$ and $x_i$. The Monte-Carlo sample operation is exactly identical to dropout \cite{b41}.\\
By given input x and the weights $w$, which can be sampled from $q(w)$, the predictive distribution of our interest is defined as:

 \begin{equation}
 \begin{split}
    p(y^*|x^*,D) \approx \int p(y^*|x^*,w)q(w)\mathrm{d}w \\ \approx \frac{1}{T}\sum_{t=1}^T p(y^*|x^*,\hat{w}_t)
 \end{split}
 \end{equation}
This is referred to as Monte Carlo dropout (MC- dropout).

\subsection{Integrated Method}
Our method intelligently decomposes the computation between edges and fog node, which fits the distributed setting and also preserves the user data privacy. The whole process is sketched as follows and the Pseudo code is shown in Algorithm 1.

\begin{itemize}
\item Firstly, $FN$ trains an initial model M using m data samples, where m is a hyperparameter that depends on the dataset. To generalize, we denote model as $M^t$, where $t$ is the number of round. We use LeNet as the model.\\

\item FN dispatches the model $M^t$ to the edge devices. For example, let's say, we have four devices, called $E_1, E_2, E_3, E_4$, and each receives $M^t$ from $FN$.\\

\item All edge devices implement Active Learning locally with \textbf{Maximal Entropy/BALD/VR} acquisition function. MC-dropout approximates variational inference. More specifically, during every acquisition (totally R acquisitions), edge devices train $M^t$ by another N data samples, chosen from data pool whose size is much larger than N.\\
\item Then, the edge devices upload the weights of models $M^t_1, M^t_2, M^t_3, M^t_4$ to the centralized $FN$.\\
\item $FN$ aggregates the weights either by averaging or choosing the best-trained model, and pass it to next round $t+1$ if necessary.
\end{itemize}

\begin{algorithm}
\caption{Algorithm}
\begin{algorithmic}
\If {t==0}
\State \text{set initial training images number} $initrain=20$ 
\State \text{form initial training set} $X_{ini}$
\State \text{train initial model} $M_{ini}$
\State \text{dispatch model} $M_{ini}$ to device $D_1,D_2,D_3,D_4$
\Else 
\For{j=1,2,3,4} \text{four devices (in Parallel)}
\For{t=1,2..T}
\For{$i= 1 \text{ to } 200$}
\State $\log p(x_i)$, $p(x_i)$ = $\text{BNN}_{j}^{t-1}$ ($x_i$)
\EndFor
\State \textbf{end}
\State $[x_1^t,x_2^t,..x_{10}^t] = \text{Acquisition}(\log p(x_i),p(x_i))$
\State $\text{BNN}_{j}^t=\text{Train}(x_1^t,x_2^t,..x_{10}^t)$
\EndFor
\State \textbf{end}
\EndFor
\State \textbf{end}
\EndIf
\State \textbf{end}
\end{algorithmic}
\end{algorithm}

T is the acquisition number, we experimented 10, 20, 30, 40 in the next section. BNN is the MC-dropout Bayesian Neural Network, Acquisition function might be Maximal Entropy, BALD or VR introduced in the second section.

In our experiments, we set $m$ equal to 20, and we only consider one round. FL \cite{b6} suggests sending the gradients back to the centralized node and the main training work is carried out on centralized node. In this case, more rounds are required. It fits the application that does not require high-level real-time reaction. In our case, we upload the parameters of the trained model to the fog node and we may also combine the two ways with the consideration of both latency and computation burden at the edge side.

The synchronization is not obligatorily required. If less devices upload in one round, the accuracy might be significantly influenced, but not necessarily if we average the weights on centralized fog node. Thus, there is no fetal problem if asynchronization happens.

\begin{figure}
\centerline{\includegraphics[width=80mm,scale=0.5]{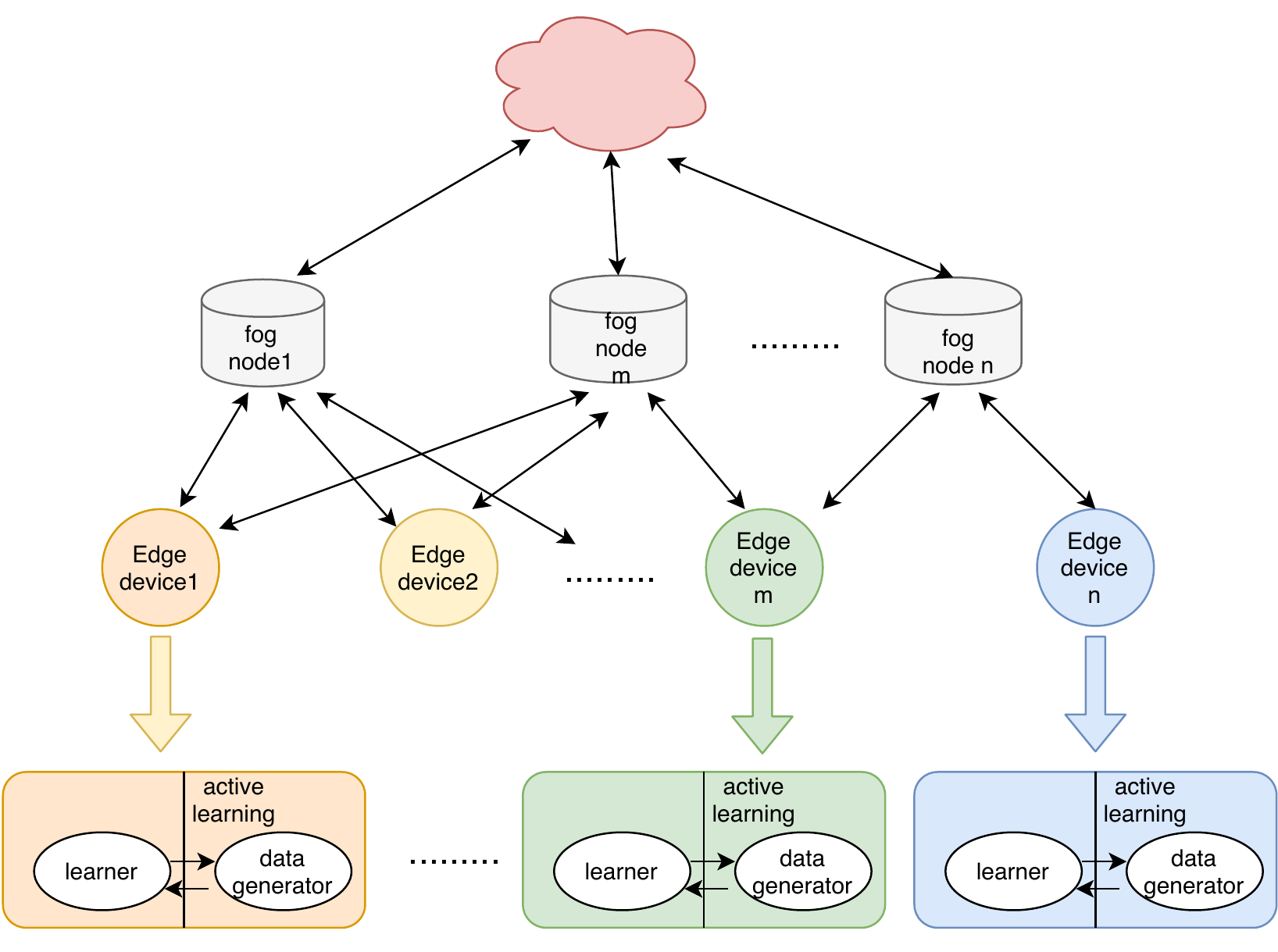}}
\caption{Overview of the scheme.}
\label{dia}
\end{figure}

Our solution that is tailored on the fog platform. It is shown in Fig. \ref{dia}, where the fog nodes work as the middleware between edge devices and cloud server. In addition, the fog node is connected to the edge devices that have the similar task to implement. For instance, a fog node might be linked to all the surveillance cameras to detect the particular object. Every camera possesses a trained model dispatched by the FN from previous round, and it keeps training by the images generated locally under Active Learning framework. It is followed by an operation that send the refined models back to fog node.

\section{Experiments and Results}\label{sec:eval}
In this section, we discuss the experimental setup along with the dataset used for our evaluation, and the results of these experiments.

\textbf{Experiment Setup:}\\
We consider two settings in our experiments: non-massive distribution and massive distribution. Firstly, we consider the non-massive case, a small number of distributed devices, let's say four edge devices and one centralized node.
Initially, we trained LeNet model by 20 images at the centralized node (Fog Node), and then dispatch the model to the edge devices. On the devices side, we further trained the model by additional data points that are generated locally. They are acquired by entropy, bald or variational ratios, and this operation will iterate several times. We will compare the performances by different acquisition strategies in the result part. Then it is followed by updating the refined models from all the devices to the centralized FN. The FN will average the parameters of the models for the next round if it is necessary.Notably, in this paper, we assume the data generated from different edge devices are from the same distribution but unbalanced. In other words, we randomly shuffle the whole training dataset, split it and distribute them to edge devices. All the sub-dataset contains 10 classes, with different proportions.

All the experiments are implemented by Python language, more specifically, Pytorch package \cite{b38} are used to build and train LeNet. The codes are run in Mac Os system (High Sierra) with version 10.13.6, with RAM 16 GB.

\textbf{Data Set:}\\
We implement the methods on MNIST dataset \cite{b8}, which is a real data set of handwritten images, containing numbers from 0 to 9, totally 10 classes. It has 60000 images for training data and 10000 for the test, . All the images have already been pre-processed, built with the size as $28*28$. It is the basic benchmark to test the classification performance of neural network in machine learning.


\subsection{Experiment \rom{1}: Effective Window Size}
\begin{figure}[htbp]
\centerline{\includegraphics[width=80mm,scale=0.5]{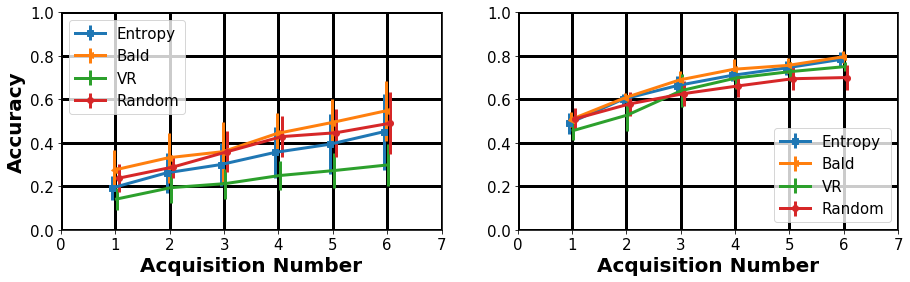}}
\caption{Learning curve: without(left)/with(right) initial training.}
\label{4}
\end{figure}

To enable the AL capability, firstly, we need a model that is able to measure the uncertainty, no matter which strategy we use, BALD, maximal entropy or VR. Namely, the model has been trained by a set of data, whose amount depends on the complexity of the data. Otherwise, AL has no significant difference with uniformly/randomly choosing data. The left one of Figure. \ref{4} shows the learning curve of the model without initial training, the rights one is the case with 20 images initial training. In the left plot, random case outperforms entropy and VR, and slightly worse than Bald. In this case, bald is the best option, but with heavier computation cost. Instead, as shown in the right plot, almost all the AL strategies outperform randomly-choosing dataset. 

\begin{figure}[htbp]
\centerline{\includegraphics[width=80mm,scale=0.5]{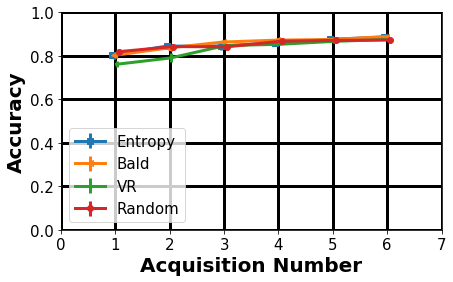}}
\caption{ Learning curve: Well-Trained Model.}
\label{6}
\end{figure}

Moreover, if we have the model that has already been well trained, which means the model has been trained to the extent that the uncertainty measurement won't help significantly comparing with uniformly picking data. It is shown in Figure \ref{6}, as we can tell there is no big difference between applying AL and randomly choosing dataset. Here we run the experiments for 5 times, we plot the error bar indicating the standard deviations.

\subsection{Experiment \rom{2}: AL acquisition number}
\begin{figure}[htbp]
\centerline{\includegraphics[width=90mm,scale=0.5]{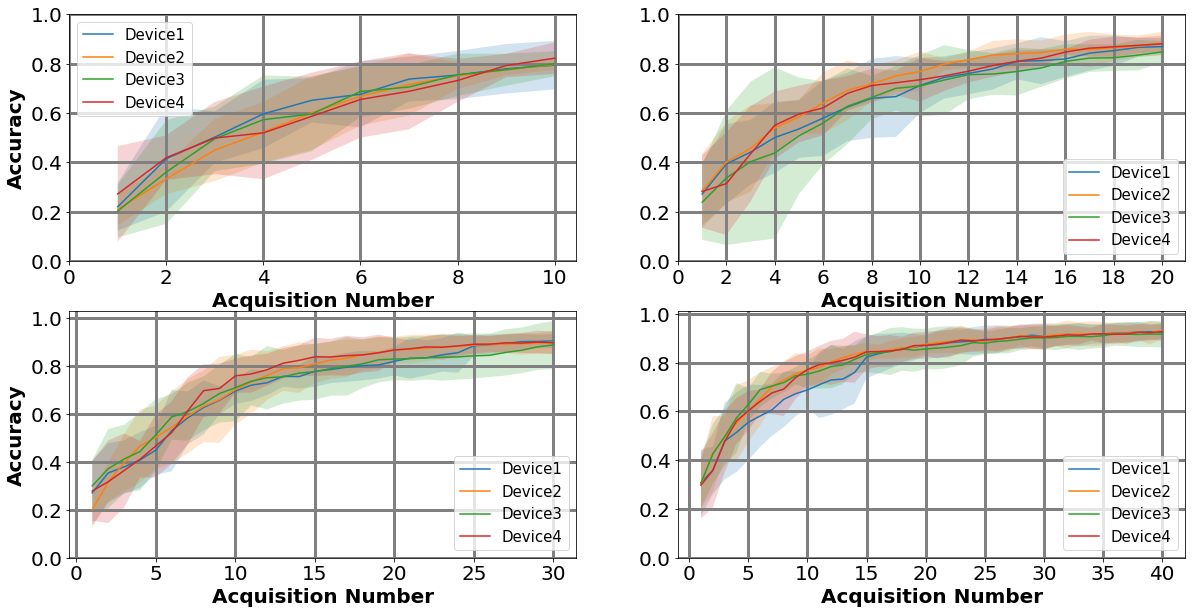}}
\caption{Learning Curve of Edge Devices for 10, 20, 30 and 40 acquisitions of data.}
\label{5}
\end{figure}

In this series of experiment, we study how does the acquisition number influence the performance. Recall that during every data acquisition, we include 10 additional images for further training. Figure~\ref{5} illustrates the learning curve of edge devices for 10, 20, 30, 40 acquisitions accordingly. Here, we run the experiments for 5 times and we also plot the standard deviations. The training on edge devices are independent; namely, with the different dataset, which conforms to the practical situation, the data is generated locally and independently.   

The first observation is that every device has its own learning curve in terms of highest accuracy it may reach and the shape of the curve as the data at every device is different though they are from the same distribution.
In addition, we build the data pool by randomly choosing 200 images from the whole dataset (10000) at every iteration of acquisition, in order to reduce the computing cost as all the data in the pool are being measured the uncertainty. It is the main reason we run the experiments several rounds to test the real performance of our method. 

In the beginning, the variance of accuracy is high ( the stripe is wider), not just due to the randomness when we build the data pool, but also the variation when we measure the uncertainty. With the increment of training data (acquisition number), the stripe becomes more and more narrow. When acquisition number is 10 and 20, four devices end up with different accuracy, while when acquisition number is 30 or 40, they are almost the same. As a summary, we suggest to choose acquisition number between 10 and 20, otherwise, we can randomly choose images.


\begin{figure}[htbp]
\centerline{\includegraphics[width=80mm,scale=0.5]{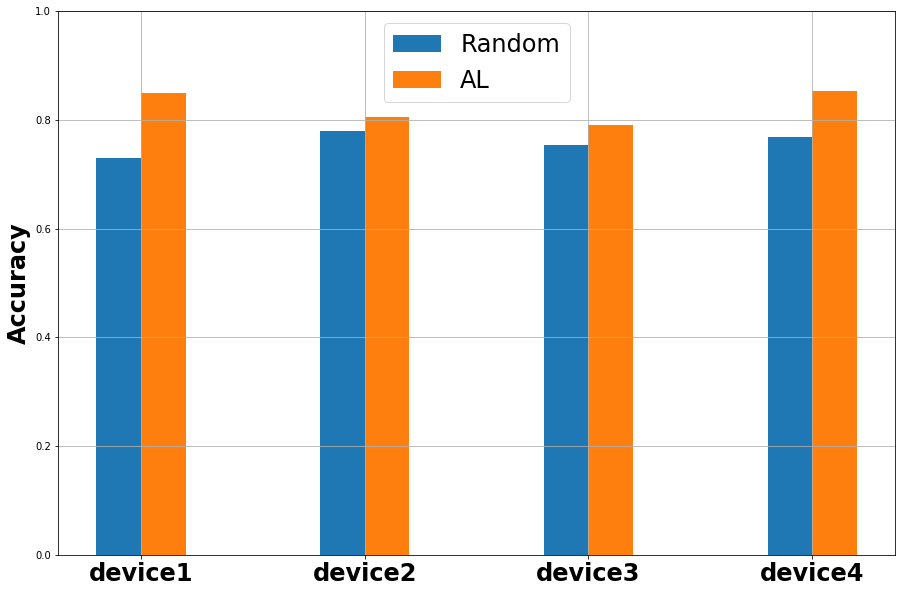}}
\caption{Active Learning Vs Random Sample (10 Acquisitions).}
\label{al_rand10}
\end{figure}

\begin{figure}[htbp]
\centerline{\includegraphics[width=80mm,scale=0.5]{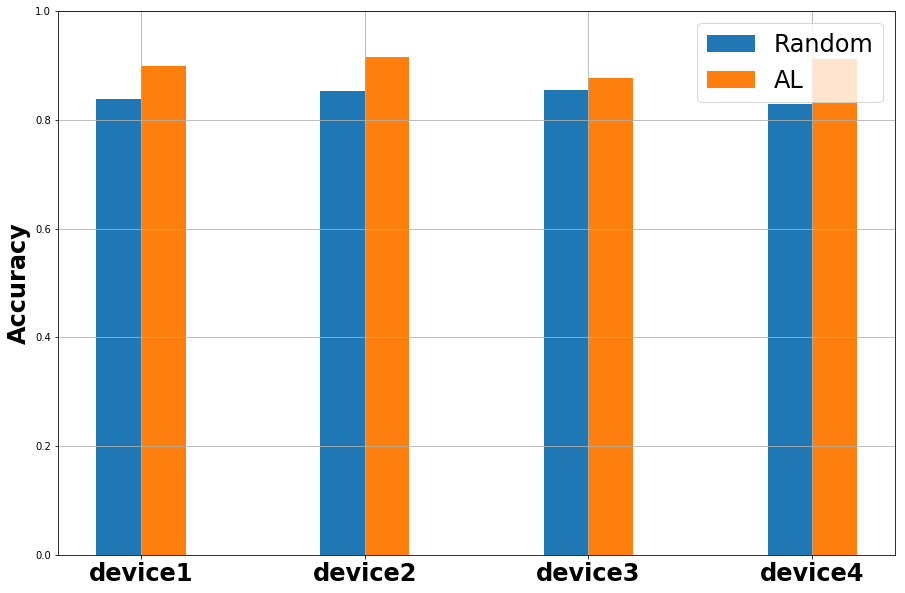}}
\caption{Active Learning Vs Random Sample (20 Acquisitions).}
\label{al_rand20}
\end{figure}

Moreover, we also plot Figure \ref{al_rand10} and \ref{al_rand20} that illustrate the superiority of AL when acquisition number is 10 and 20 and the model is initially trained by 20 images. As we have already shown there is no big difference between Bald and entropy if we have initial training in Figure \ref{4}, we use entropy strategies for the sake of computation cost.

\subsection{Experiment \rom{3}: Centralized vs Decentralized}
The performance of fog node, which is aggregated from the distributed nodes depends on the way of aggregation: choosing the optimal model, averaging the parameters of models from diverse devices or stacking the weights by decomposition of the model. Heuristically, when training data size is small, the accuracy between devices varies to the considerable extent, thus, picking the optimal model leads a higher accuracy than averaging parameters. While, averaging strategy has the high robustness. Our method mainly shows its strength when the dataset is small. Instead, when the training size is large, though, the learning curves are not necessary the same, they end up with the similar performance. 

Distributed configuration can also be interpreted as the stochastic training case, every (big) batch is trained on individual device. It might bypass being trapped in the local optimal when we train all the dataset in one place.

We compare the performance difference between centralized computing and decentralized computing (our method). More specifically, we compare the accuracy on FN by applying our approach, with the result obtained by training a dataset with the size triple bigger (4*N) than on every edge device (N) as we have four edge devices in our experiment. For instance, if we train the model by 100 data points on the edge device, then we compare it with the result directly training 400 data samples on FN. The details are shown in Table ~\ref{tab1}, and the columns indicate the different number of acquisitions from data pool, during every acquisition we pick ten images, Acq 10 means the model is further trained by another 100 images on every edge device. To compare, we directly trained the initial model with 400 images since we have four devices, every device is trained by 100 images. And then we compared the accuracy with two aggregation strategies: average and optimal model. Note that it is arguable that how many images we train directly on FN to compare since the model is not directly trained by 400 images when we apply FL, we train the model by 100 images on every device. Nevertheless, here we train the model by training data with the size equal to the number locally trained on every device time the number of devices, considering the worst case.\\

\begin{table}[htbp]
\caption{Fog Node Performance with/without Federated Learning}
\begin{center}
\begin{tabular}{|c|c|c|c|c|}
\hline
 &\textbf{Acq 10} & \textbf{Acq 20} & \textbf{Acq 30} & \textbf{Acq 40}\\
\hline
\textbf{\makecell{FN without FL \\(No. of train data)}} & 400 & 800 &1200 & 1600\\
\hline
\textbf{\makecell{FN with FL \\(No. of train data)}} & 100 & 200&300&400\\
\hline
\textbf{\makecell{Accuracy\\ (FN without FL)}}& 0.73 & 0.882& 0.811 &0.901\\
\hline
\textbf{\makecell{Accuracy\\ (FN with FL (ave))}}&0.8& 0.909 &0.881&0.918 \\
\hline
\textbf{\makecell{Accuracy\\ (FN with FL (opt))}} &0.854& 0.915&0.944&0.963\\
\hline
\end{tabular}
\label{tab1}
\end{center}
\end{table}

\begin{figure*}
  \includegraphics[width=18cm,height=7cm]{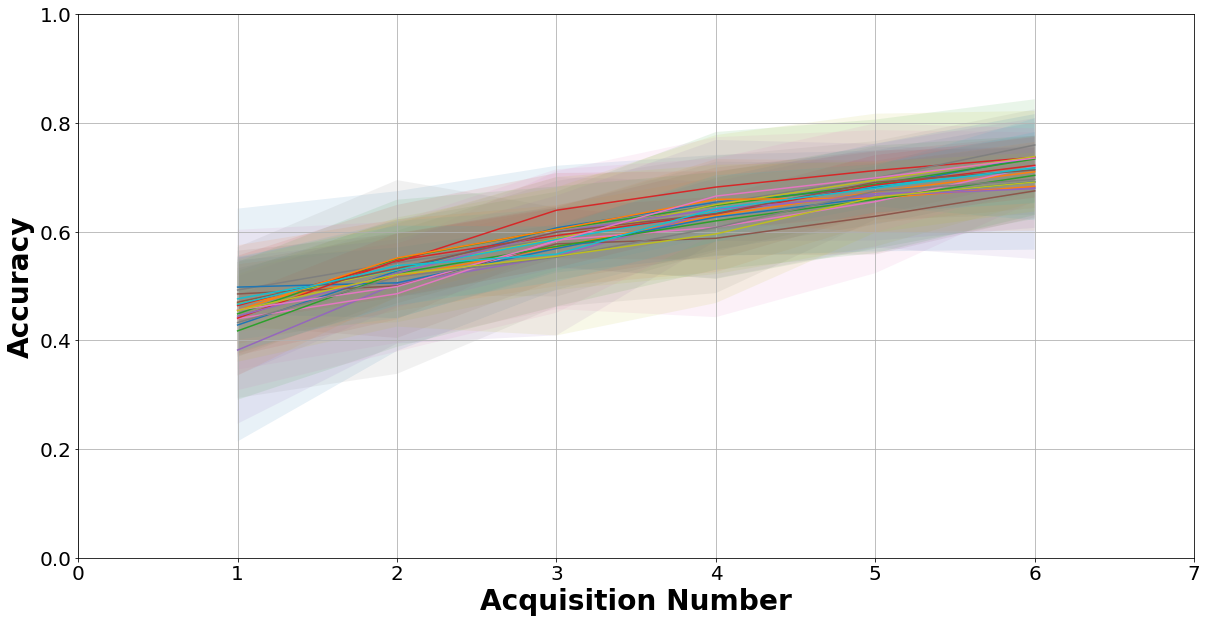}
  \caption{learning curves: 20 devices, trained by 60 images.}
  \label{20devices}
\end{figure*}
\subsection{Massive Distribution}
\begin{figure}[htbp]
\centerline{\includegraphics[width=70mm,scale=0.5]{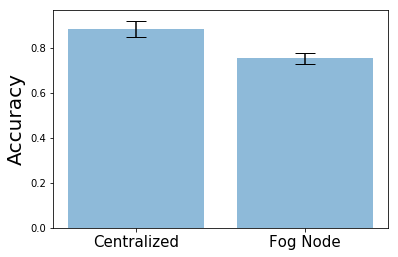}}
\caption{Accuracy from the centralized fog node where we have 20 devices where each devices is trained by 60 images, comparing with centralized node trained by 1200 images.}
\label{acc_com}
\end{figure}

\begin{figure}[htbp]
\centerline{\includegraphics[width=70mm,scale=0.8]{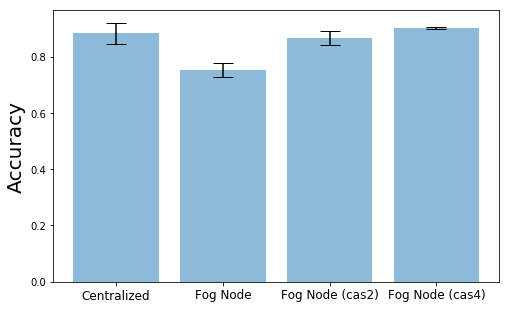}}
\caption{Accuracy from the centralized fog node where we have 20 devices where each devices is trained by 60 images, comparing with centralized node trained by 1200 images and cascade with closest one neighbouring node and three neighbouring nodes.}
\label{acc_com_all}
\end{figure}

\begin{figure}[htbp]
\centerline{\includegraphics[width=80mm,scale=0.8]{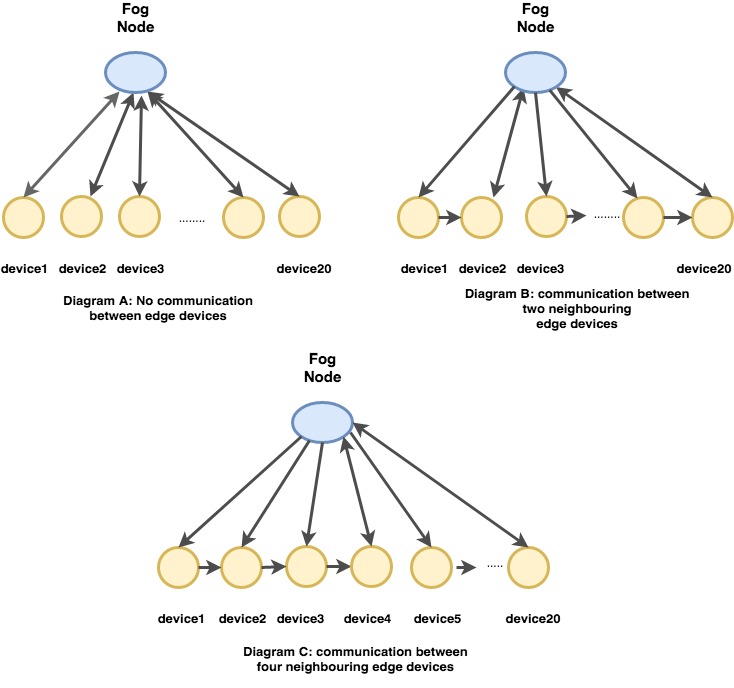}}
\caption{Architecture of massively distributed setting. Diagram A indicates no communication between devices, which only communicate with fog node. Diagram B shows the communication only between two devices, and diagram C shows the communication between four devices.}
\label{arch_com}
\end{figure}

When the number of edge devices is large, the advantage of AL is not as obvious as the case (4 edge devices). Assuming with 1200 data images, if we have 4 edge devices, every device is trained by 300 images, while if we have 20 devices, then every device 'sees' only 60 images. The learning curve of the second case is demonstrated in Figure \ref{20devices}. In the second case, the centralized device that uses the ensembling model works worse than one centralized machine trained directly by 1200 images ($60<<1200$), as shown in Figure \ref{acc_com}. The centralized one ends up with the accuracy 0.89, while the distributed case is 0.75. It can be solved by enabling the communication between devices, cascading the training process, namely, after one device completes the training process, shares the trained model with the close neighbouring device. Doing so, the computation will slow down as one device has to wait for the neighbour node, but the accuracy will be improved. In Figure \ref{acc_com_all}, we demonstrate the case when two edge devices and four edge devices are cascaded, with the accuracy 0.87 and 0.9, with 2 times and 4 times speed slow down accordingly. The configuration is shown in Figure \ref{arch_com}, where we include the architecture that there is no communication between edge devices and the communication between two devices and four devices. In the cascading case, there exists the dependence between edge devices.

\section{Conclusion}\label{sec:Conc}
Industrial Internet is possible on the Fog Computing Platform that integrates all the necessary requirements. In this paper, we for the first time discussed Active Learning in a distributed setting tailored to the Fog platform consisting of distributed edge devices and a centralized fog node. We implemented active learning in edge devices to down scale the necessary training set and reduce the label cost. Also, we discussed the window size as a critical factor that decides the effectiveness of AL in the result section. The first criteria is that we need to train the initial model to the extent that it may roughly measure the uncertainty, otherwise the model disable to choose more representative data. The second rule is that we shouldn't over-train the model, otherwise there is no significant difference between AL and randomly choosing data. The specific size depends on the complexity of the dataset. We explored two configurations: massive distribution, non-massive distribution and gave the corresponding solution. In non-massive setting, we disable the communication between edge devices. In massive setting, for the sake of accuracy, we suggest the communication between edge devices, cascading the model training procedure, with the cost of slowing down the training procedure. We presented evidence that it performs similarly to centralized computing with a reduced communication overhead, latency and harvesting the potential privacy benefits. 

In the future, we will apply the idea on the IoT application and test the performance. In addition, we will study the additional acquisition
functions and also address the privacy issues in more details.



\end{document}